# Electrically Tunable Optical Absorption in a Graphene-based Salisbury Screen

Vrinda Thareja, Juhyung Kang, Hongtao Yuan, Kaveh M. Milaninia, Harold Y. Hwang,  Yi Cui, Mark L. Brongersma*,

Geballe Laboratory for Advanced Materials, Stanford University, Stanford, CA 94305

* Correspondence should be addressed to: brongersma@stanford.edu

**We demonstrate a graphene-based Salisbury screen consisting of a single layer of graphene placed in close proximity to a gold back reflector. The light absorption in the screen can be actively tuned by electrically gating the carrier density in the graphene layer with an ionic liquid/gel. The screen was designed to achieve maximum absorption at a target wavelength of 3.2 μm by using a 600 nm-thick, non-absorbing silica spacer layer. Spectroscopic reflectance measurements were performed *in-situ* as a function of gate bias. The changes in the reflectance spectra were analyzed using a Fresnel based transfer matrix model in which graphene was treated as an infinitesimally thin sheet with conductivity given by the Kubo formula. Temporal coupled mode theory was employed to analyze and intuitively understand the observed absorption changes in the Salisbury screen. We achieved ~ 6 % change in the optical absorption of graphene by tuning the applied gate bias from 0.8 V to 2.6 V, where 0.8 V corresponds to graphene's charge neutrality point.**

Graphene, a two-dimensional (2D) sheet of hexagonally arranged carbon atoms, has gained significant attention for its unique electrical, optical and mechanical properties.[1–4] These include its ability to absorb 2.3 % of the incident light over a broad range of wavelengths, an extremely high charge carrier mobility on the order of ~200,000 cm$^2$/Vs for suspended, high-quality graphene and an extraordinary intrinsic mechanical strength.[5-6] This makes graphene a promising candidate for many applications such as ultrafast photodetectors, modulators, solar cells and transparent electrodes.[7–11] Whereas the single-layer-absorption is impressive, it is not strong enough for many optoelectronic applications as virtually all the light is transmitted. Moreover, for active devices such as electro-optical modulators and actively controlled thermal emitters, dynamic control over absorption is required as well.[12] In this work, we explore an electrically-tunable Salisbury screen device capable of achieving both increased absorption and active absorption modulation in a single graphene sheet.

Recently, there have been a number of excellent efforts directed at enhancing the optical absorption in graphene. For example, plasmonic nanostructures have been used to locally concentrate fields in this 2D material with the aim to increase the efficiency of graphene-based photodetectors.[13,14] Theoretically, complete light absorption in graphene has also been predicted under critical-coupling conditions by patterning it.[15] Integrating graphene with an optical waveguide has been employed to achieve a longer light-graphene interaction length.[16] Graphene microcavity photodetectors have also been demonstrated that produce increased absorption in graphene due to enhanced optical fields inside a high quality factor (Q) resonant cavity.[17]

Here, we demonstrate a graphene-based Salisbury screen capable of boosting the absorption of infrared light in a graphene layer well beyond that of a suspended sheet. Furthermore, we show that we can actively tune the absorption in the graphene layer by electrical gating with an ionic gel. The original Salisbury[18] screen was invented by the American engineer Winfield Salisbury in 1952 and was employed in some of the first radar absorbent materials.[19] In its original implementation, a weakly absorbing layer of graphite (i.e. multi-layered graphene) was spaced from a metallic back reflector by a transparent spacer layer. It was shown that (for some target wavelength) the net reflection from the system can be made to vanish and unity absorption can be achieved. To first order, this occurs when the direct reflection from the absorbing layer and the light returning from the back reflector are equal in magnitude and opposite in sign. For this to happen, the spacer layer thickness ($t$) has to be approximately equal to a quarter of the wavelength of the incident light in the spacer layer with refractive index $n_{Spacer}$, i.e. $t \approx \lambda/4n_{Spacer}$. In a second-order theory, multiple reflections between the screen and the reflector have to be accounted for.

When the mirror is a perfect electrical conductor (PEC) featuring a very high electrical conductivity, the spacer thickness should exactly equal one quarter of the targeted operation wavelength to maximize absorption. However, real metals at optical frequencies allow for some field penetration (up to a skin depth) and this gives rise to a reflection phase pick up that deviates from that of a perfect electrical conductor. As a result, a thinner spacer layer thickness can be used to achieve the destructive interference condition. The field penetration comes at the cost of some optical losses in the metal. However, when a very strongly absorbing film like graphene is used in conjunction with a real, but highly conductive metal reflector, it is possible that most of the absorption takes place within the film as opposed to the metal. As a result, an active tuning of the absorption properties of the film can have a significant impact on the reflected signal. In this work, we combine experiment and theory to analyze the nature and magnitude of the absorption changes one can achieve with a graphene-based Salisbury screen device.

Figure 1a shows a schematic of our proposed device configuration. It consists of a graphene sheet that is judiciously-spaced from a metallic back reflector by a transparent 600-nm-thick silicon dioxide ($SiO_2$) spacer layer. A 200-nm-thick gold (Au) film was used as the metal back reflector due to its high reflectance in the mid-infrared (IR) part of the electromagnetic spectrum. The entire stack was deposited on a silicon (Si) substrate by standard electron beam (e-beam) evaporation. The $SiO_2$ layer was produced by high density plasma enhanced chemical vapor deposition (HDPECVD). The thickness of this spacer layer was chosen such that an incident wave at a target wavelength of ~3.2 μm produces a standing wave with a maximum in the electric field intensity at the location of the graphene sheet (Fig. 1b). This choice facilitates both strong light absorption and a relative insensitivity of the absorption to the exact position of the screen. The mid-IR spectral range around 3 μm is of great importance for many applications, including IR vibrational spectroscopy and thermal radiation control. Commercially available CVD graphene (from ACS Material) was transferred onto the $SiO_2$ spacer. Using e-beam evaporation, large gold pads on the order of ~ 1 cm x 0.5 cm were then deposited to serve as source and drain contacts for electrical transport measurements on graphene. A larger, laterally-placed Au pad of approximately 2 cm x 1 cm was also deposited to electrically gate and control the carrier density of the graphene sheet with an ionic gel. This was followed by spin-coating a smooth, 4.5 μm thick layer of the ionic gel, derived from the ionic liquid DEME-TFSI [N,N-diethyl-N-methyl-N-(2-methoxyethyl) ammonium bis(trifluoromethylsulfonyl) imide]. The ion gel has a high refractive index that is comparable to that of $SiO_2$ ($n_{gel}$ = 1.43).[20] As such it forms a low Q optical cavity together with the silica layer that facilitates resonant recirculation of the incident wave. For this reason, the presence of the ion gel layer can further enhance the light absorption in the graphene layer over a conventional screen that would have air above the conductive sheet. In this way the gel plays an important optical role in addition to its electrical role of gating the device.

The variation of the drain-source current through the graphene channel with the applied gate bias across the ion gel is shown in Fig. 1c. The measurements were performed using a semiconductor parameter analyzer (Agilent B1500). A constant drain-source voltage ($V_{DS}$) of 10 mV was applied while monitoring the resultant drain-source current ($I_{DS}$) as a function of the applied gate bias ($V_G$). Upon application of a gate bias, ionic charges accumulate at the ion gel/graphene and ion gel/gold pad interfaces.[21] This in turn causes the carrier density and hence the conductivity of graphene to change. A minimum in $I_{DS}$ occurs at a gate bias referred to as the Dirac voltage ($V_D$). This condition corresponds to the charge neutrality point (CNP) in graphene.[22] At the CNP, the Fermi level of graphene passes through the point of intersection of the conduction and valence band. The occurrence of the CNP at a positive $V_G$ (= 0.8 V) indicates that the graphene sheet in our devices was effectively p-doped with the Fermi level lying in the valence band as shown schematically in the band diagram in Fig. 1c. The magnitude of $I_{DS}$ increases again for voltages above the CNP due to accumulation of electrons in the conduction band, making graphene effectively n-type.

The electrical data in Fig. 1c will be used to predict the bias-dependent reflection properties of the Salisbury screen. To this end, we first describe an equivalent circuit model for our device that links the carrier density/position of the Fermi level to the applied gate voltage. The conductivity σ across the spectral range of interest can in turn be calculated based on the position of the Fermi level using the theoretical formalism developed by Kubo (discussed later). The optical absorption in the structure is proportional to the conductivity and the square of the electric field magnitude at the location of the graphene sheet. An optical (transfer matrix) model is finally constructed to calculate the field distribution and the optical reflection properties for the screen.

The electric double layers (EDLs) formed at the ion gel/graphene and ion gel/gold (Au) pad interfaces serve as capacitor structures with an approximate spacing between the "plates" of a few nanometers.[23] Physically, this effective distance is linked to the spacing of the ions in the gel and the mobile carriers in graphene or Au. We define a total double layer capacitance ($C_{EDL}$) as the geometric capacitance per unit area of the EDLs formed by the series combination of the Au/ion gel interfacial capacitance and graphene/ion gel interfacial capacitance. At first, one might think that these are the only capacitors at play in the system. However, due to the finite density of states of 2D conductors like graphene, such materials do not behave like the metallic plates of a conventional capacitor structure. Here, one needs to account for the substantial impact of band-filling/band-emptying upon charging/discharging of the 2D graphene 'plate' that features a small density-of-states compared to regular metals. To construct useful circuit models for such unusual capacitor plate materials, Luryi introduced the concept of a quantum capacitance ($C_Q$).[24] He found that one can take into account the band-filling/band-emptying with a second quantum capacitance placed in series to the usual geometric capacitance of a parallel plate capacitor.[25] As a result, the applied gate voltage creates both an electrostatic potential difference $\phi$ across the geometric capacitance and produces a shift in the Fermi level $E_F$ that is linked to the quantum capacitance of graphene:[26]

$$|V_G - V_D| = \phi + \frac{E_F}{e} , \quad (1)$$

In this equation $e$ is the electron charge. The magnitude of electrostatic potential $\phi = ne/C_{EDL}$ has a linear dependence on the charge density $n$ similar to that for a regular metal-plate capacitor. On the other hand, the magnitude of $E_F$ is proportional to the square root of the charge density $n$:[27]

$$E_F = \hbar v_F \sqrt{\pi n} , \quad (2)$$

Here, $v_F$ is the Fermi velocity and $\hbar$ is the reduced Planck's constant. Taking $C_{EDL} = 1.45 \times 10^{-3}$ F/m$^2$ as an example, which is a reasonable estimate based on the dielectric properties of the ion gel and the thickness of the double layer (few nanometers), we can plot the estimated charge density $n$ vs. effective gate bias $V_G$-$V_D$ (Inset of Fig. 1c). It should be noted that these types of estimates of the double layer capacitance tend to be first-order estimates based on the not so well-defined nature of this capacitor structure. With knowledge of the carrier density and Eq. (1), one can calculate the dependence of $E_F$ on the gate voltage.

The changes in $\sigma$ that result from bias-induced movements in the Fermi energy level of graphene can be calculated using the Kubo formula:[28,29]

$$\sigma = -\frac{ie^2}{4\pi\hbar} \ln \frac{2E_F - (\hbar\omega - 2i\hbar\Gamma)}{2E_F + (\hbar\omega - 2i\hbar\Gamma)} - \frac{ie^2 k_B T}{\pi\hbar(\hbar\omega - 2i\hbar\Gamma)} \left[ \frac{E_F}{k_B T} + 2\ln\left(e^{\frac{-E_F}{k_B T}} + 1\right) \right] , \quad (3)$$

where $\omega$ is the angular frequency, $\Gamma$ is the scattering rate for the mobile carriers, $k_B$ is the Boltzmann constant and $T$ is the temperature. The dependence of conductivity of a graphene sheet on wavelength is plotted in Fig. 1d for four different possible locations of the Fermi level: 0 eV (CNP point), 0.13 eV, 0.16 eV, 0.195 eV. These values of $E_F$ are achieved at 0.8 V (CNP point), 1.8 V, 2 V and 2.6 V respectively. The spectral dependence of $\sigma$ can be understood by realizing that its magnitude is governed by two major contributions that dominate in different wavelength regimes. At longer wavelengths, the dominant contribution to the conductivity comes from the free-carrier response (i.e. intraband excitations). This contribution is described fairly well by the Drude model. At shorter wavelengths, the optical response is dominated by interband transitions and is independent of wavelength.[30] As a result, the conductivity approaches the well-known universal value of 60 µS at the shorter wavelengths.[31] It can be seen that an increase in the bias voltages causes the conductivity to decrease over the entire wavelength range of interest. The exact functional dependence of the conductivity on the applied bias is however different at each wavelength as each of the conductivity curves shows a 'swing' from a high to a low value at a photon energy $E_{ph} \approx 2 E_F$. This swing is caused by the fact that a minimum amount of photon energy $E_{ph} \approx 2 E_F$ is required to stimulate an interband transition in graphene (as dictated by the state filling that is schematically depicted in the two band diagrams in Fig. 1d).

In order to theoretically predict the optical response of our graphene-based Salisbury screen in the infrared, a transfer matrix model [32] was used, wherein graphene was treated as an infinitesimally thin sheet with a conductivity given by Eq. (3). The calculated spectral reflectance curves for this structure at the different Fermi level positions/values of the gate bias are shown in Fig. 2. Each of the reflectance curves shows series of four Fabry-Perot-like oscillations. The magnitude of the free spectral range of the oscillations indicates that the structure serves as an asymmetric Fabry-Perot resonator, where the resonator thickness is equal to the sum of $SiO_2$ and ion gel layers. To confirm this point, we also show the reflectance for the same structure with a semi-infinite ion gel layer (dotted line in Fig. 2). Here, the oscillations have disappeared as the reflective ion gel/air boundary is removed. Instead, a single reflection minimum is observed at a wavelength of ~3.2 μm. At this wavelength, the electric field shows a maximum value at the location of the graphene sheet (See Fig. 1b). Application of an electrical bias primarily changes the depth of the reflection peaks and does not significantly affect the spectral position. Similar to the trends for graphene's conductivity (Fig. 1d), a gradual increase in the Fermi level first impacts the spectral reflectance at long wavelengths and then at shorter ones.

Next, we measured the dependence of the optical response of the fabricated graphene-based Salisbury absorber on gate bias. A Fourier Transform Infra-Red (FTIR) spectrometer was employed to measure the reflectance spectra. For practical applications, we are particularly interested in quantifying the ability to modulate the reflectance. To this end, we will analyze the normalized differential reflectance $\frac{R_{E_F} - R_D}{R_D}$, which is proportional to the difference between the reflectance at a given Fermi energy level of interest (or applied gate bias) $R_{E_F}$ and the reflectance at the CNP given by $R_D$. Based on $I_{DS} - V_G$ measurements discussed earlier (Fig. 1c), the CNP for our devices was achieved at an applied bias of 0.8 V. The dots in Fig. 3a show measurements of the normalized differential reflectance at gate biases of 0.8 V, 1.8 V, 2 V and 2.6 V. The solid lines in Fig. 3a represent the calculated differential reflectance spectra based on the transfer matrix model at the corresponding Fermi energy levels in graphene of 0 eV (CNP), 0.13 eV, 0.16 eV and 0.195 eV (using $C_{EDL} = 1.45 \times 10^{-3}$ F/m$^2$ as previously mentioned). Good agreement is obtained between experimental and theoretical spectra. The curve corresponding to 0.8 V ($E_F = 0$ eV) coincides with the horizontal axis since the reflectance at this gate voltage is used as the reference for normalization. The four peaks in this figure occur at the spectral locations where the calculated reflectance spectra exhibited the four Fabry-Perot resonances (Fig. 2).

Figure 3a shows that the largest change in reflectance due to a Fermi level change (0 eV to 0.195 eV) in graphene occurs at a wavelength of 4.17 µm and not at the Salisbury resonance wavelength of ~3.2 µm. This can be understood by noting that the change in reflectance with the applied gate bias can be expressed as a product of two contributions. One of them is optical in nature and is the change in reflectance for a given change in graphene's conductivity. The other is electrical in nature and quantifies the change in conductivity with a change in the applied voltage:

$$\frac{\Delta R}{\Delta V} = \left(\frac{\Delta R}{\Delta \sigma}\right)\left(\frac{\Delta \sigma}{\Delta V}\right) \quad , \tag{4}$$

The observed spectral dependence of the bias-induced reflectance changes can be understood from the spectral dependence of each of the two terms on the right side of Eq. (4). The changes $\left(\frac{\Delta R}{\Delta \sigma}\right)$ can be calculated from the transfer matrix model. Figure 3b shows how the reflectance changes as the conductivity of the absorber layer is varied over a wide range of conductivity values. The plot highlights that the screen could have achieved unity absorption if only graphene had conductivity as high as 3 mS. The inset shows that for typical conductivity values of graphene (0 µS to 100 µS), an almost linear dependence of the reflectance on conductivity is observed, with only a slight dependence on probe wavelengths. Fig. 3c plots the magnitude of $\left(\frac{\Delta R}{\Delta \sigma}\right)$ versus the illumination wavelength. This quantity is largest when the illumination wavelength meets the Salisbury screen condition *($\lambda \approx 4tn_{Spacer}$)*. This is expected as for this wavelength ($\lambda = 3.2$ µm) the field is maximized at the location of the screen (Fig. 1b). Figure 1b also shows the field profiles for the other Fabry-Perot resonance wavelengths of 2.7 µm and 4.2 µm, which show similar standing wave amplitudes of the electric field but lower amplitudes of the field at the location of the graphene sheet. Figure 1d showed that the largest changes in the conductivity with increasing voltage $\left(\frac{\Delta \sigma}{\Delta V}\right)$ occur when the magnitude of $E_F$ reaches half the photon energy $E_{ph}$. As a result, the changes for low bias voltages are largest at the long wavelength side of the spectrum and this causes the overall changes in the reflectance for the largest bias voltage of 2.6 V to be highest at 4.2 µm instead of at the Salisbury screen condition: $\lambda = 3.2$ µm. These changes amount to about 3.3% per volt. That said, in going from 2 V to 2.6 V (i.e. at high bias) the largest changes in the reflectance do occur at $\lambda = 3.2$ µm.

To better understand the observed reflection changes for our graphene-based Salisbury screen, we employ coupled mode theory.[33] Coupled mode theory effectively hides some of the device complexity and thereby often affords a simple, basic intuition about the device operation. Following this approach, we treat our Salisbury screen as an asymmetric Fabry-Perot resonator as depicted schematically in Fig. 4a. We consider a wave of amplitude $s_+$, with frequency $\omega$, incident on the resonator that can excite a resonator mode with amplitude $a$:

$$\frac{da}{dt} = i\omega_o a - \left(\frac{1}{\tau_i} + \frac{1}{\tau_e}\right) a + \kappa s_+ \quad , \tag{4}$$

where $\kappa$ quantifies the degree of coupling between the resonator and the incident wave $s_+$. The magnitudes of $1/\tau_i$ and $1/\tau_e$ represent the internal and external amplitude decay rates respectively and $\omega_o$ is the resonant (angular) frequency. Assuming a steady state situation ($da/dt = 0$), one can use Eq. (4) to calculate the reflection coefficient for the structure:

$$r = \frac{(1/\tau_e) - (1/\tau_i) - i(\omega - \omega_o)}{(1/\tau_e) + (1/\tau_i) + i(\omega - \omega_o)} \quad , \tag{5}$$

and hence find an expression for the reflectance:

$$R = \frac{\left(\frac{1}{\tau_e} - \frac{1}{\tau_i}\right)^2 + (\omega - \omega_o)^2}{\left(\frac{1}{\tau_e} + \frac{1}{\tau_i}\right)^2 + (\omega - \omega_o)^2} \quad , \tag{6}$$

Each amplitude decay rate can be related to an optical quality factor:

$$Q = \frac{\omega_o \tau}{2} \quad , \tag{7}$$

and the total quality factor $Q_{\text{tot}}$ of the system can be found by inversely adding the relevant Q factors associated with different loss processes. The magnitude of $Q_{\text{tot}}$ measures the ratio of the stored energy in the cavity, $E_{\text{stored}}$ to the total power dissipated $P_{\text{tot}}$ from/in the cavity:

$$Q_{\text{tot}} = \frac{\omega_o * E_{\text{stored}}}{P_{\text{tot}}} \quad . \tag{8}$$

In our model, graphene and gold are treated as dispersive media in the frequency range of our interest (IR). The silica and ion gel materials are assumed to have a constant permittivity in the IR. For a medium with constant permittivity, the average stored energy is defined as:

$$E_{\text{stored}} = \frac{\epsilon_o}{4} \int \epsilon |E|^2 dV \quad , \tag{9}$$

where $\epsilon_o$ is the permittivity of free space, $\epsilon$ is the relative permittivity of the medium and $E$ is the electric field amplitude in the medium. The electric field intensity is integrated over volume $V$ of the medium of relative permittivity $\epsilon$. Gold is a dispersive medium and follows the Drude model in the considered spectral range.[34] The average energy stored for dispersive media obeying Drude model is given by:[35]

$$E_{\text{stored}} = \frac{\epsilon_o}{4} \int \left[ \epsilon' + \frac{2\omega \epsilon''}{\Gamma} \right] |E|^2 dV \qquad , \qquad (10)$$

where $\epsilon'$ is the real part of relative permittivity, $\epsilon''$ is the imaginary part of the relative permittivity of the medium. The power dissipation (for both dispersive and dispersion-less media[36]) is given by:

$$P_{\text{loss}} = \frac{\epsilon_o}{2} \omega \epsilon'' \int |E|^2 dV \qquad , \qquad (11)$$

Based on the coupled mode theory analysis, $Q_{\text{tot}}$ has contributions from both internal and external loss processes. The internal losses result from power absorbed in both graphene ($P_{\text{gr}}$) and the gold back reflector ($P_{\text{Au}}$). The external Q arises due to the optical power that leaks out of the cavity $P_{\text{rad}}$. The total power loss in Eq. (8) is given by:

$$P_{\text{tot}} = P_{\text{rad}} + P_{\text{gr}} + P_{\text{Au}} \qquad , \qquad (12)$$

The total energy stored in the cavity is the sum of the energy stored in each layer of the resonator i.e.:

$$E_{\text{tot}} = E_{\text{gr}} + E_{\text{Au}} + E_{\text{silica}} + E_{\text{gel}} \qquad , \qquad (13)$$

where $E_{\text{gr}}$, $E_{\text{Au}}$, $E_{\text{silica}}$, $E_{\text{gel}}$ are the stored energy within graphene, gold, silica and ion gel layers respectively.

Figure 4b shows the value of $Q_{\text{tot}}$ at the four considered resonance wavelengths and for the four different values of $E_F$ considered above. The magnitude of this quantity was derived from the line-widths of the various resonances. For all the values of $E_F$, virtually the same decrease in $Q_{\text{tot}}$ from about 12 to 5 is observed on going from the shortest (2.67 μm) to the longest wavelength (5.81μm) resonance. This change of about a factor of two is related to the approximate doubling of the wavelength since Q ~ $\omega_0$ per Eq. (8). The independence of the trend on the location of $E_F$ is explained by the fact that the magnitude of $Q_{\text{tot}}$ is largely controlled by the external (i.e. leakage) loss.

Due to the high conductivity of the Au film in this spectral range, the magnitude of the power loss is very small compared to the radiation loss and loss in the graphene. As such, the electrically induced changes in the reflectance are effectively controlled by the changes in the loss within graphene. To understand the impact of the losses in graphene on $Q_{\text{tot}}$, we next analyze the electrically-induced changes on the quality factor of graphene $Q_{\text{gr}}$. Figure 4c shows the dependence of $Q_{\text{gr}}$ for the same resonance wavelengths and magnitudes of $E_F$ as in panel 4b. For $E_F = 0$eV, the magnitude of $Q_{\text{gr}}$ again decreases by about a factor 2 from 340 to 160, which is reasonable given the virtually constant conductivity (i.e. absorption properties of graphene). Its value is significantly larger than $Q_{\text{tot}}$, indicating that the optical losses in the graphene layer only contribute a small fraction to the total loss. This results in a relatively simple linear dependence of the reflectance from our screen at small values of σ (Inset to Fig 3b). When a voltage is applied to raise the Fermi level in graphene, the conductivity of graphene is lowered (Fig. 1d), causing a rise in the magnitude of $Q_{gr}$. The largest increase in $Q_{gr}$ from 309 to 1123 occurs at the wavelength of 4.17 µm corresponding to a maximum conductivity change. Based on our simulations this corresponds to an absorption change in graphene of ~6 %.

In conclusion, we have demonstrated a graphene-based Salisbury absorber that employs the Salisbury screen concept to enhance the absorption in graphene. The absorption in this planar absorber can be tuned by using a gate voltage. The reflectance changes within the absorber as a function of Fermi energy levels in graphene were modeled by a transfer matrix model and the experimental findings match well with the theoretical predictions. To relate the reflectance changes to the absorption within graphene and to understand the absorption mechanism, we made use of the temporal coupled mode theory. This theory explains the magnitude of the reflectance changes and why there is a simple linear dependence of the reflectance on the electrically-induced changes in graphene's conductivity. We achieved changes in the reflectance of up to ~3.3 % per volt with our design by varying the Fermi level in graphene. This shows the benefits of placing graphene in a Salisbury screen type configuration.

**Author Information**
Corresponding Author: Mark L. Brongersma, Brongersma@stanford.edu
* Geballe Laboratory for Advanced Materials, 476 Lomita Mall, Stanford CA, 94305

**Author Contributions**
The manuscript was written through contributions of all authors. All authors have given approval to the final version of the manuscript.

**Acknowledgement**
V.T., J.K., K.M.M., and M.L.B acknowledge support from the Department of Energy Grant no. DEFG07ER46426. "H.T.Y., H.Y.H. and Y.C. acknowledge the support from the Department of Energy, Office of Basic Energy Sciences, Division of Materials Sciences and Engineering, under contract DE-AC02-76SF00515."

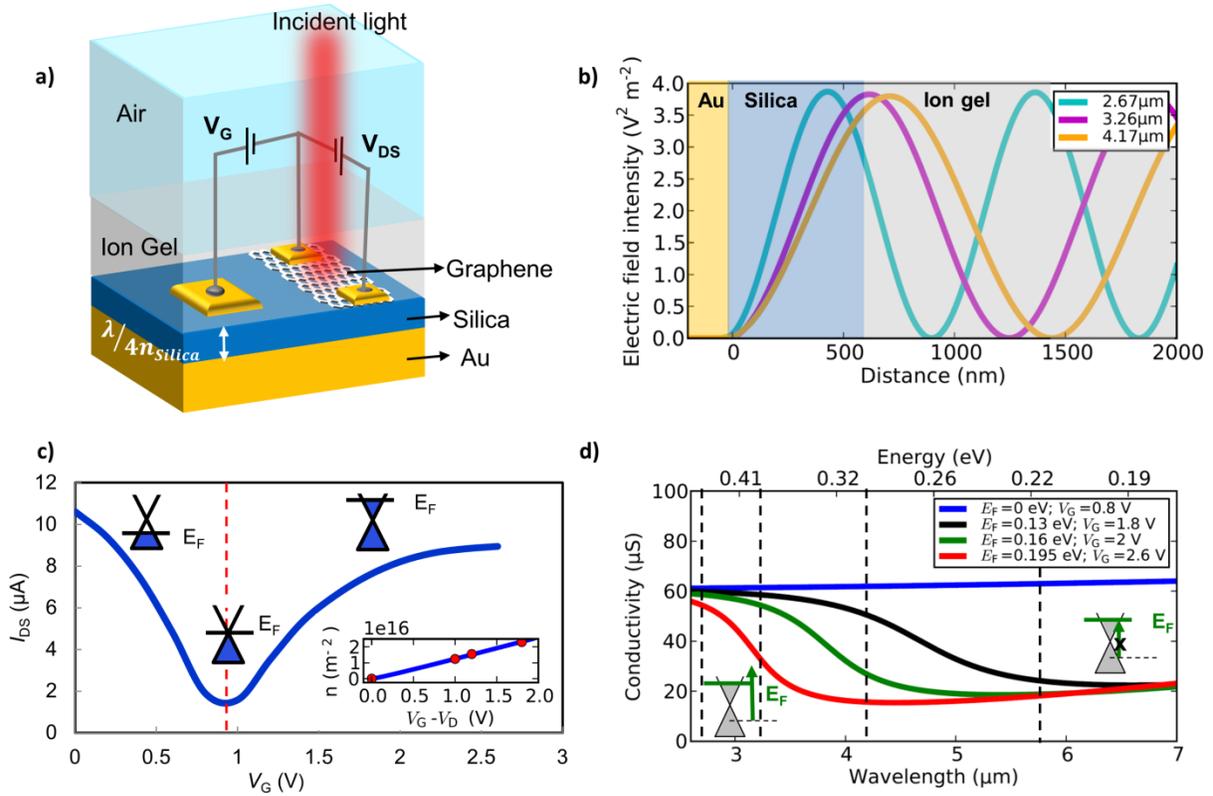

**Figure 1. (a)** Schematic of the graphene-based Salisbury absorber consisting of a stack with air/ion gel/silica/gold. **(b)** Electric field intensity profile within the device structure for illumination wavelengths of 2.7 μm, 3.2 μm and 4.2 μm. The color convention for the different device layers matches those in panel (a). **(c)** Experimentally found dependence of the drain-source current ($I_{DS}$) on gate voltage ($V_G$). The Dirac cones indicate the locations of the Fermi energy levels at different applied gate voltages. Inset – charge carrier density and Fermi level variation with effective applied gate bias. **(d)** Calculated spectral dependence of the conductivity of graphene in the mid-infrared spectral range for different Fermi energy level positions/magnitude of the gate bias.

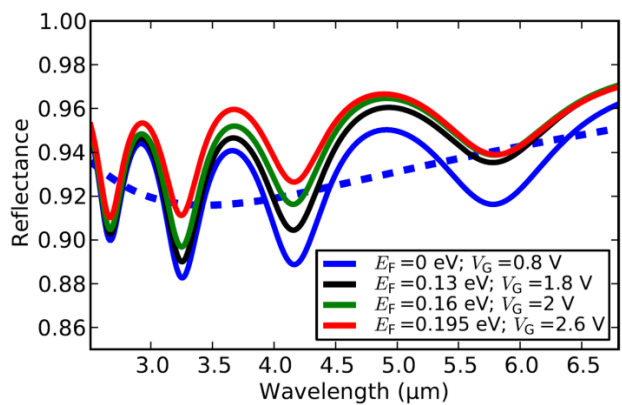

**Figure 2.** Reflection spectra for the Salisbury screen at different positions of the Fermi energy level in graphene. The dotted blue curve represents the reflectance for a structure with semi-infinite ion gel instead.

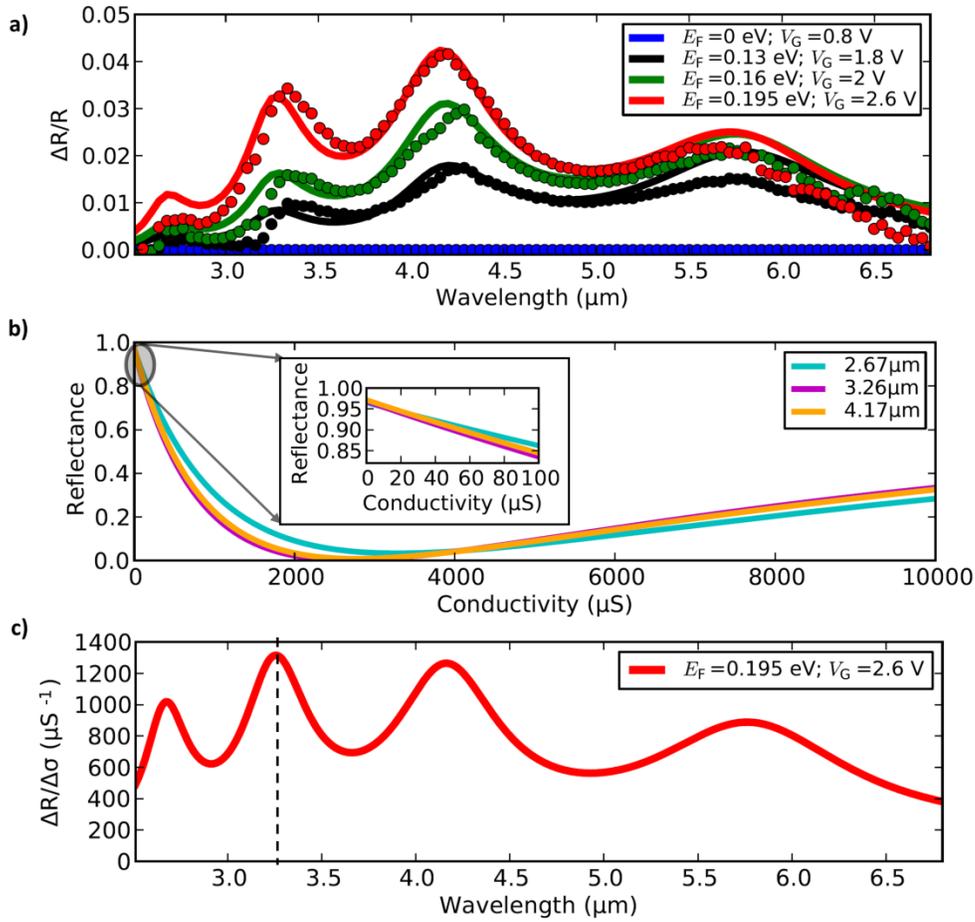

**Figure 3. (a)** Normalized differential reflectance spectra for different Fermi energy levels in graphene. The dotted lines are the experimental measured values and the solid lines are transfer-matrix model-based vales. **(b)** Reflectance from an absorbing layer of variable conductivity. Inset – Zoomed in reflectance for conductivity values in the range of 0μS to 100μS. **(c)** Differential change in the optical reflectance from the Salisbury screen with the conductivity of graphene as a function of wavelength.

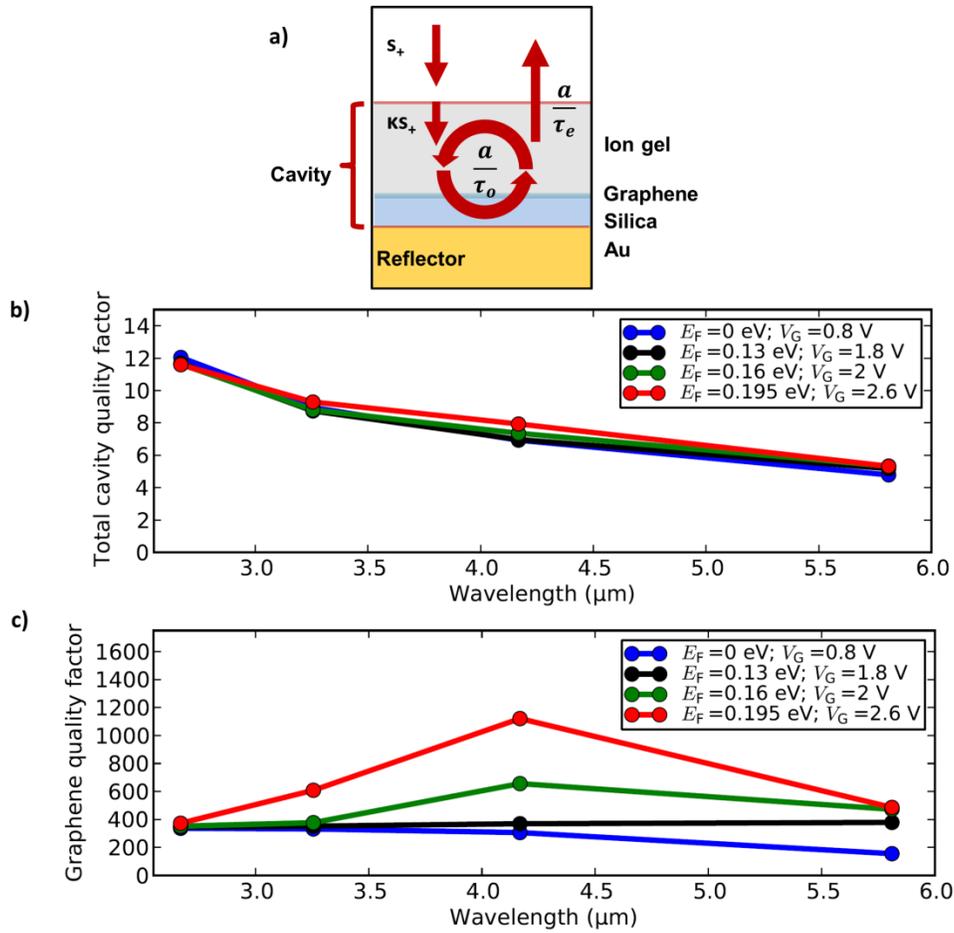

**Figure 4**. **(a)** Coupled mode theory schematic explanation for equation (4) . **(b)** Total cavity quality factor spectra variation with Fermi energy levels in graphene. **(c)** Quality factor spectra in graphene for different Fermi levels in graphene